\documentclass[a4paper,11pt]{article}
\pdfoutput=1 

\usepackage{jheppub} 
\usepackage[]{algorithmic}
\usepackage[ruled,vlined]{algorithm2e}
\usepackage[T1]{fontenc} 
\usepackage{graphics}
\usepackage{multirow}

\newcommand{\concat}{%
  \mathbin{{+}\mspace{-8mu}{+}}%
}

\title{Fast Simulation of a High Granularity Calorimeter by Generative Adversarial Networks}

\author[a,b,1]{Gul Rukh Khattak,\note{Corresponding author.}}
\author[a]{Sofia Vallecorsa,}
\author[a]{Federico Carminati,}
\author[b]{Gul Muhammad Khan}

\affiliation[a]{CERN,\\1211 Geneva 23, Switzerland}
\affiliation[b]{University of Engineering and Technology Peshawar,\\Jamrud Road Peshawar Khyber Pakhtunkhwa, Pakistan}

\emailAdd{gul.rukh.khattak@cern.ch}
\emailAdd{sofia.vallecorsa@cern.ch}
\emailAdd{federico.carminati@cern.ch}
\emailAdd{gk502@uetpeshawar.edu.pk}

\abstract{We present the 3DGAN for the simulation of a future high granularity calorimeter output as three-dimensional images. We prove the efficacy of Generative Adversarial Networks (GANs) for generating scientific data while retaining a high level of accuracy for diverse metrics across a large range of input variables. We demonstrate a successful application of the transfer learning concept: we train the network to simulate showers for electrons from a reduced range of primary energies, we then train further for a five times larger range (the model could not train for the larger range directly). The same concept is extended to generate showers for other particles (photons and neutral pions) depositing most of their energies in electromagnetic interactions. In addition, the generation of charged pion showers is also explored, a more accurate effort would require additional data from other detectors not included in the scope of the current work. Our further contribution is a demonstration of using GAN-generated data for a practical application. We train a third-party network using GAN-generated data and prove that the response is similar to a network trained with data from the Monte Carlo simulation.

The showers generated by GAN present accuracy within $10\%$ of Monte Carlo for a diverse range of physics features, with three orders of magnitude speedup. The speedup for both the training and inference can be further enhanced by distributed training.}

\begin{document}
\maketitle{}
\section{Introduction}
\label{sec:intro}

Particles undergo complex stochastic interactions upon contact with materials. The modeling of these interactions is further complicated by the large number of secondary particles involved. The Monte Carlo simulation depends on repeated random sampling to produce a snapshot of the particle interactions with a detector. Simulation is crucial in most High Energy Physics (HEP) experiments and is extremely resource-intensive. More than $50\%$ of the current computing resources of the HEP community are utilized in simulation alone~\cite{hsf2019}. In the future that need will increase further due to the higher luminosity and granularity of future experiments, and it will not be possible to create a corresponding increase in computing resources. The main motivation for the fast simulation is to incorporate other faster alternatives to decrease the cost of future experiments.  Current fast simulation approaches are mostly based on parametrization~\cite{Altfast, gflash, nomad} or lookup table~\cite{frozenShower} approaches, providing between 10 and 100 times speedup while achieving different levels of accuracy. 

Generative Adversarial Network (GAN)~\cite{2014Goodfellow} is a training paradigm for deep generational neural networks. Other approaches for generative networks include Variational autoencoders~\citep{2014autoencoding} and autoregressive models~\citep{2016pixelrnn}, etc. All of these approaches have their own strengths and weaknesses. The autoencoder-based methods often produce blurry images while the pixel-based methods are not only slow to evaluate but also suffer from limited capacity. The GAN approach has been able to demonstrate highly realistic and sharp images as compared to other approaches~\cite{2014Goodfellow}. There have been many recent variants of the GAN methodology, such as WGAN~\cite{wgan}, StackGAN~\cite{stackgan}, and Progressive GAN~\cite{progressiveGAN} further enhancing the quality and resolution of generated images. The generative problem often involves intractable probability densities and thus methods involving likelihood estimates are not practical for most purposes. GAN can learn a distribution implicitly since it does not rely on the explicit computation of probability densities. Therefore the GAN approach is suitable for the generation of a wider range of data such as musical notes~\citep{musicGAN}, natural language~\citep{langGAN}, medical data~\citep{medicalGAN, medicalGAN2}, natural scenes~\citep{2014Goodfellow}, faces~\citep{progressiveGAN} and image denoising\citep{denoiseGAN}. Since simulation is essentially a generative problem thus a successful image generation model can be exploited in this domain. A potential advantage of deep generative models is that the generated distribution does not have to be explicitly defined, and thus even the real data from a detector can be simulated directly.

We leverage the GAN methodology to generate HEP calorimeter output. Calorimeters are special HEP detectors that record particles through the measurement of the energies deposited by them. These detectors can be regarded as huge cameras taking pictures of particle interactions. The Monte Carlo simulation for the detector output is extremely precise but highly expensive both in regards to the simulation time and resources. For most HEP experiments the calorimeter is a simulation bottleneck, consuming, for example, more than $80\%$ of the simulation time for the ATLAS experiment~\cite{Altfast}. We generate the calorimeter cells as monochromatic pixelated images with the cell energy depositions as our pixel intensities. 

The 3DGAN~\cite{icip18_3dgan, 3dgan_icmla, Belayneh2019} is the first effort where the detector output is generated retaining correlations in all three spatial dimensions. We employ three-dimensional convolutional layers in our network, while previously detector output was generated either as a two-dimensional image or a concatenated set of two-dimensional images. We demonstrate our approach for a high granularity detector with higher spatial resolution and thus consequently much larger image dimensions than previous such efforts. 
We pre-process the cell depositions by taking a power less than one, thus decreasing the dynamic range of corresponding pixel intensities and improving the convergence. We employ a multi-step training process to generate images, from a complex multivariate distribution, for a large range of input conditions. We also perform extensive validations from diverse viewpoints including vision and deep learning, as well as, physics-based evaluation. The network scores highly on all the platforms both for the pertinence to the training data and for maintaining sufficient diversity. The details of the network development and the validation from different perspectives have been presented previously~\cite{3dgan_icmla}. The current work is geared more towards the physics community and thus is limited to only physics-based validation. Previously we simulated electrons coming with energies from a wide spectrum by employing our multi-step training. Now we successfully extend the same approach to simulate additional particle types such as photon and neutral pion where most of the energy is lost in electromagnetic interactions. We perform an additional investigation to prove that the GAN could accurately reproduce the signature features of a particular particle type. We also undertake some preliminary exploration of the charged pion simulation and generation of rare events. We finally present a successful practical example of using the GAN-generated data in a typical reconstruction tool, demonstrating that the GAN-generated images could provide similar performance as Monte Carlo images.

The current paper is organized as follows. Section~\ref{sec:prev_work} presents the overview of past efforts for fast simulation of HEP calorimeters exploiting deep neural networks. The next section (Section~\ref{sec:calorimeter}) describes the Monte Carlo training dataset. The basic structure of the calorimeter and the important features of our data are discussed. 
Section~\ref{sec:3dgan} describes how the GAN approach is adapted to the problem of HEP detector simulation. The approach has been exploited for the generation of particles with predominantly electromagnetic showers. The results for comparison to Monte Carlo simulation are presented in Section~\ref{sec:results}. Some preliminary work is also carried out for the charged pion simulation as presented in Section~\ref{sec:chpions}. Another study exploring the simulation of rare modes is presented in Section~\ref{sec:rare}. Section~\ref{sec:gan_training} presents a practical application for the use of GAN-generated data. Finally, Section~\ref{sec:conc} summarises the main contributions and presents some future suggestions.

\section{Previous Work}
\label{sec:prev_work}
Fast simulation is already incorporated in existing experiments through approaches like parametrization~\cite{nomad, gflash, Altfast, fastcalosim} and Lookup tables~\cite{frozenShower}, etc. Usually a part of the simulation is replaced by fastsim where some tradeoff between speed and accuracy can be feasible. Following the same concept deep learning has also been explored to generate simulation data. Fast simulation using neural networks can be regarded as a special type of parametrization, with the weights of the neural network as parameters, optimized through a training process. 

\begin{table}[]
    \centering
    \caption{Calorimeter simulation with deep learning}
    \begin{tabular}{p{0.2cm} p{3.5cm} p{1.4cm} p{2cm} p{2.6cm} p{2.4cm}}
         \\
         No.& Model & Algorithm & Architecture &Condition & Output \\
         \hline
         \hline
         \\
         1& LAGAN~\cite{de2017learning} (2017) & GAN & 2D Locally connected& Particle Type as discrete labels& $25\times25$\\
         2& CALOGAN~\cite{paganini2017calogan} (2017) & GAN &2D Locally connected& $E_P\sim{} U(1, 100)$ GeV& layer1:~$3\times96$ layer2:~$12\times12$ layer3:~$12\times6$\\
         3& 3DGAN initial prototype~\cite{icip18_3dgan} (2018) & ACGAN & Conv3D & $E_P\sim{} U(2, 500)$ GeV & $25\times25\times25 $\\
         4& ATLAS~\cite{Salamani2018} (2018) & WGAN and VAE & Dense & $E_P\sim{} U(1, 260)$  GeV & vector of 266 cells.\\
         5& LHCb~\cite{gan_lhcb} (2019) & WGAN& Conv2D & Five variables related to position and momentum & $30\times30$\\
         6& HGCAL~\cite{precise} (2019) & WGAN& Conv2D and Locally connected& $E_P$ and initial impact position (x,y) & Concatenation of 7 ($12\times15$) layers\\ 
         7& 3DGAN~\cite{Belayneh2019} (2019) & ACGAN & Conv3D & $E_P\sim{} U(2, 500)$ GeV and $\theta\sim{} U(60^{\circ}, 120^{\circ})$& $51\times51\times25 $\\
         8& DijetGAN~\cite{Dijet} (2020) & WGAN& Conv2D & & vector of 7 jet variables.\\
         9& ILD~\cite{high2020} (2021) & GAN, WGAN and BIB-AE& Conv3D & $E_P\sim{} U(10, 100)$ GeV)& $30\times30\times30$.\\
         \\
         \hline
         \hline
    \end{tabular}
    
    \label{tab:sim}
\end{table}{}

The GAN technique is an unsupervised training methodology. The power of GAN lies in the fact that the target distribution does not have to be tractable and instead the training relies on a Minimax game between a discriminator (D) network and a generator (G) network. The discriminator is trained to differentiate between the target and the generated distributions, while the generator is trained to confuse the discriminator. Both the networks compete with each other till the generator manages to completely confuse the discriminator, given enough capacity for both models. At this point, the GAN is said to have converged. 

Calorimeter data have been simulated through deep generative networks in a number of recent approaches as presented in Table~\ref{tab:sim}. LAGAN~\cite{de2017learning} was one of the first fast simulation approaches based on deep learning. A simplified calorimeter was simulated as 2D jet images for high energy W bosons (signal) and generic quark/gluon jets (background). CALOGAN~\cite{paganini2017calogan} employed the LAGAN architecture to generate sets of three two-dimensional images that were then concatenated to obtain the output for a three-layered simplified calorimeter conditioned on the primary particle energy ($E_P$) ranging from $1-100$ GeV. Since then, there have been other demonstrations employing deep learning for HEP calorimeter simulation. Deep learning has been used for fast simulation of the ATLAS calorimeter~\cite{Salamani2018}. Showers with energies $1-260$ GeV and pseudorapidity $|\eta|$ in the range of $0.2-0.25$ were generated as flattened arrays of pixels, by a dense network employing both VAE and GAN methodologies. The images were also conditioned on the primary particle energy and constrained on the total energy deposition. The GAN-generated showers were reported to have better performance as compared to the VAE generated showers. WGAN has also been used to simulate the LHC detector output collapsed to a two-dimensional array of cells~\cite{gan_lhcb}. A simplified version of HGCAL was simulated as seven 2D images concatenated together, conditioned on the primary particle energy and impact position~\cite{precise}. The DijetGAN~\cite{Dijet} employed GAN for the simulation of diject events: a background process for important physics studies at LHC. A more recent approach~\cite{high2020} experimented with several GAN architectures for the simulation of high granularity calorimeter for the ILD~\cite{ild} detector as $30\times30\times30$ three-dimensional images. The simulation was limited to the orthogonally incident photons coming with $10-100$ GeV primary energy. 

The 3DGAN initial prototype~\cite{icip18_3dgan} exploited 3D convolutional networks to simulate the response of a high granularity calorimeter as $25\times25\times25$ image. The GAN setup was used to train the network for a simplified scenario involving only orthogonally incident electrons. The approach was then extended to condition $51\times51\times25$ images on both the particle energy and incident angle~\cite{Belayneh2019, 3dgan_icmla}. The more complex distribution could be generated through multi-step training, architecture, and loss function modifications (details in \cite{access2021}). We now simulate the detector output for all the particle types available in the dataset and further validate the results. The 3DGAN greatly surpasses existing efforts in the granularity and dimensions of the generated images, conditioned on both the incident particle angle and energy from a wide range, and validated in great detail from diverse viewpoints. Finally, a culmination of the effort is to test the GAN-generated data for a practical use case. 

\section{Calorimeter Dataset}
\label{sec:calorimeter}

We present a solution for the needs of future experiments with higher demands for computing resources due to increased luminosity and granularity. We, therefore, select the proposed Linear Collider Detector (LCD), designed in the context of the future Compact Linear Collider (CLIC)~\cite{CLIC} accelerator for our study. The dataset employs the GEANT4 toolkit~\citep{geant4} for the generation of the simulation data for several particle types (i.e., electrons $e$, photons $\gamma$, neutral pions $\pi^0$ and charged pions $\pi$) and is publicly available on Zenodo at \url{https://zenodo.org/communities/mpp-hep}.

\subsection{Detector Geometry}
\label{sec:geo}
\begin{figure}
    \centering
    \includegraphics[scale=0.4, trim={1cm 2cm 1cm 0cm}, clip=true]{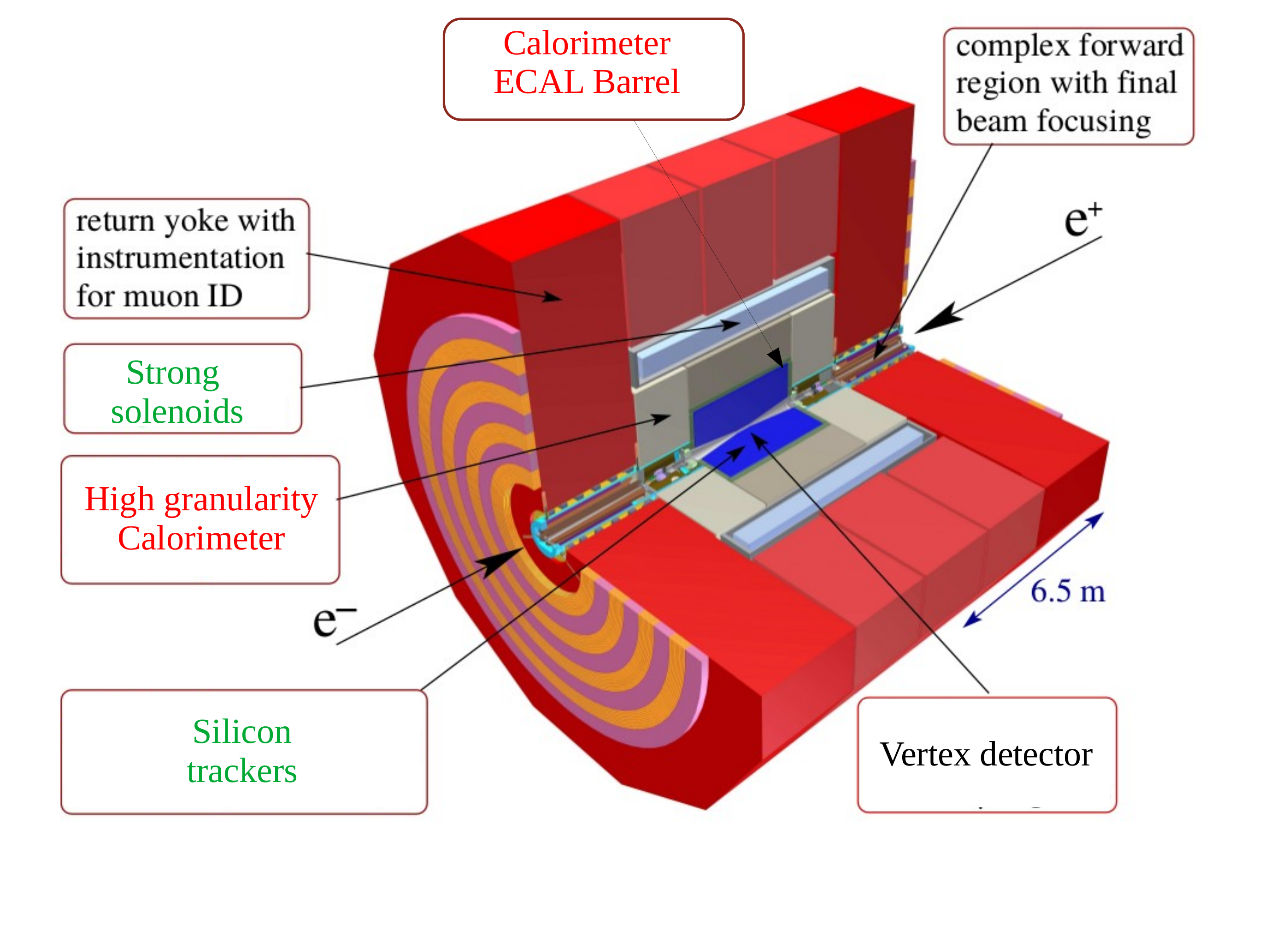}
    \caption{Schematic diagram for the CLIC calorimeter~\citep{clic_report}}
    \label{fig:clic}
\end{figure}

Figure~\ref{fig:clic} shows the proposed detector design, highlighting the main detector concepts. The basic design consideration for improving the jet energy resolution is to resolve the energy depositions of the individual particles in a jet, through a high cell granularity and precise time information. The electron and positron will collide in the central region. The trackers are shown in blue. The surrounding grey region will comprise the calorimeters. The calorimeter will be highly segmented with an electromagnetic (ECAL) and a hadronic (HCAL)  calorimeter. 

The data used for the current work is that of the ECAL central barrel region. This region has a cylindrical shape with an inner radius of $1.5~m$ with $25$ concentric layers. The proposed granularity for the ECAL cells is  $5.1\times5.1~mm^2$. The cells are arranged in the form of $25$ cylindrical layers with silicon sensor planes (active), alternating with tungsten absorber planes (passive). The simulation is carried out considering the entire detector geometry, including the material in front of the calorimeter, and the effect of the solenoid magnetic field. 

{}

\subsection{Data Features}
\label{sec:gen}
\begin{figure}
    \centering
    \includegraphics[scale=0.4, trim={0.1cm 7cm 0.5cm 0.5cm}, clip=true]{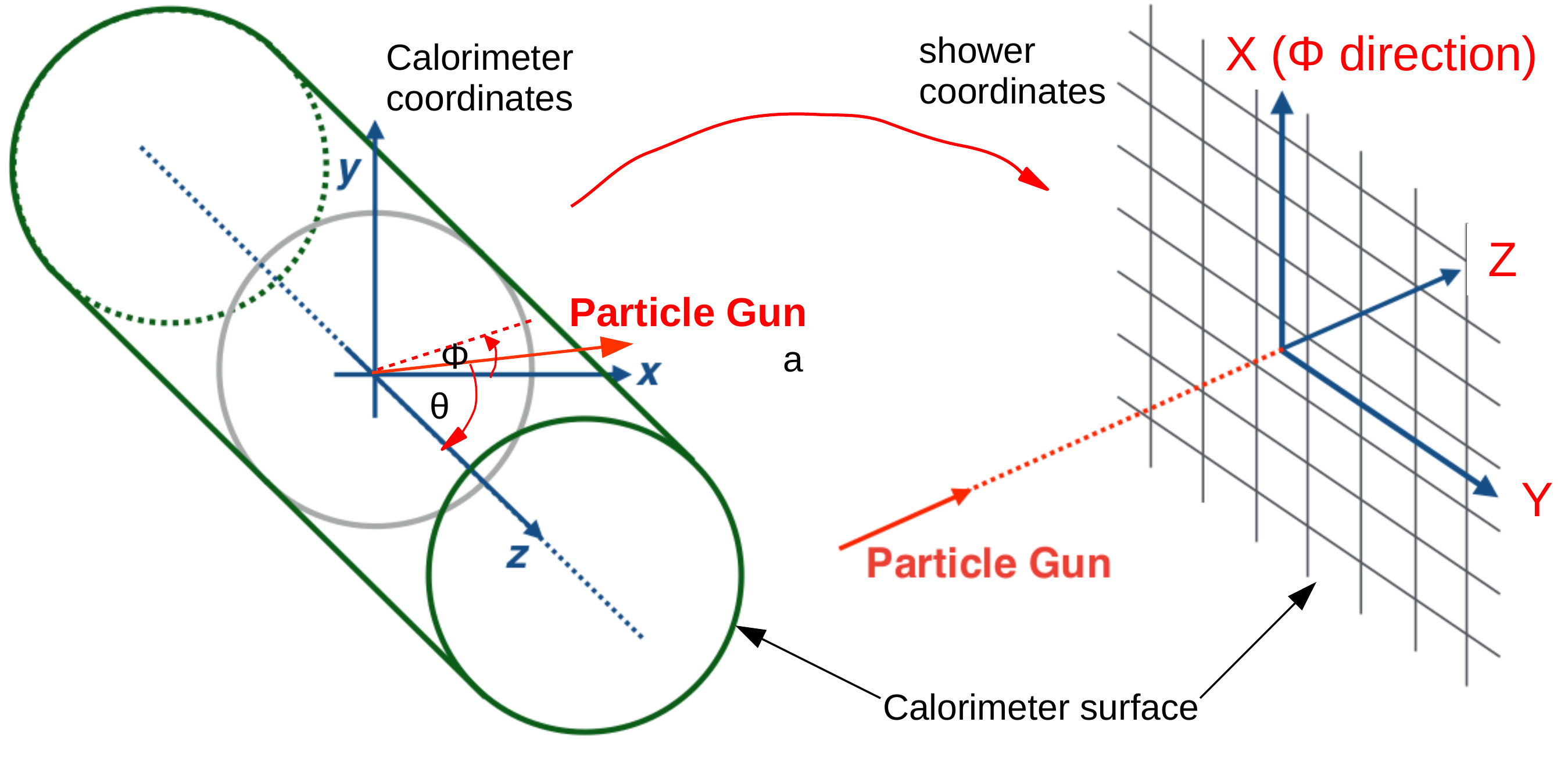}
    \caption{The calorimeter barycenter computation was done in global coordinates (left) and the cells were saved based on the local coordinates (right).}
    \label{fig:lcd}
\end{figure}{}

The energy deposits in the calorimeter cells result from the interaction of an incoming primary particle with the calorimeter material. These deposits form a characteristic shape that can be termed as an "event" or "shower". A slice around the barycenter of each shower is saved as a 3D array of energy depositions. The slicing is carried out by taking a projection of all the deposited energy on the ECAL inner surface. The barycenter of this 2D image and the point of origin of the incoming particle are then used to compute the polar angles ($\theta$ and $\phi$) corresponding to each shower. Due to the different $\phi$ granularity for each depth layer in the ECAL, a multiplicative transformation is also applied to scale every layer to look like the innermost ECAL layer. Finally, the data is saved in the HDF5 format. Each entry in the dataset comprises the 3D array with cell energy deposits, the incoming particle energy $E_P$, and the incidence angles $\theta$ and $\phi$. The energies for the incoming particles are uniformly distributed from $2$ to $500$ GeV and an incident angle ($\theta$) uniformly distributed from $60^{\circ}$ ($1.047$ radians) to $120^{\circ}$ ($2.094$ radians). 

Figure~\ref{fig:lcd} shows the particle gun position with respect to the calorimeter surface. In the global coordinate system, the $Z$ axis lies along the axis of the calorimeter cylinder. While in the local coordinates of the shower it is perpendicular to the calorimeter surface. Similarly, other axes are also transformed to the local coordinates of each sample. The $Z$ axis of our 3D images lies along with the detector depth and $X$, $Y$ are the transverse axes. The calorimeter is isotropic in the $\phi$ direction, therefore only the $\theta$ direction is incorporated for the current work. The $\theta$ value is recomputed as a weighted mean of the angles computed using the barycenter of the event and the barycenters of the $XY$ planes for each position along $Z$ (weighted by the position along $Z$).

\section{3DGAN}
\label{sec:3dgan}

The HEP simulation depends on a set of variables that impact the underlying physics processes described by the simulation. Therefore, 3DGAN uses the $E_{P}$ and $\theta$ of a particle striking the calorimeter surface as inputs, to generate the appropriate detector response. In order to provide feedback on the correspondence between generated showers and input conditions, we exploit the concept of auxiliary tasks~\cite{acgan2} together with domain-related constraints. The current works mainly describe the final optimized version of the 3DGAN model while additional details about the development process can be obtained from~\cite{access2021}. The 3DGAN is implemented using Keras 2.2.4~\cite{keras} deep learning python library with Tensorflow 1.14.0~\cite{tensorflow2015-whitepaper} as a backend. The code is available at \url{ https://github.com/svalleco/3Dgan}. 




\subsection{Pre-processing}
\label{sec:preproc}

\begin{figure}
\centering
\includegraphics[scale=0.4, trim={1cm 1cm 2cm 0.5cm}, clip=true]{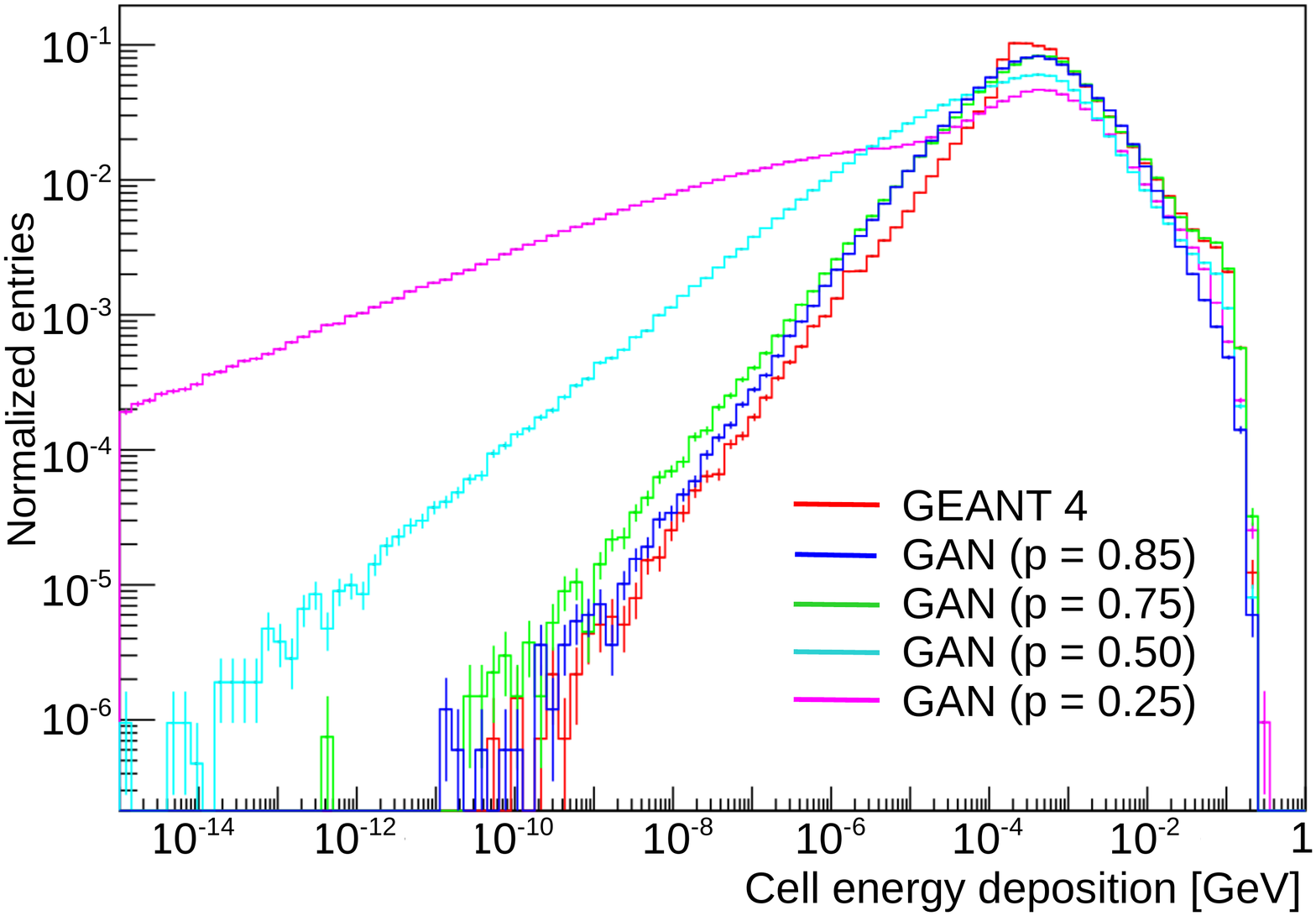}
\caption{The cell energy distribution for GEANT4 MC events (red) and GAN generated events after a pre-processing by taking the power of pixel intensities: power =$0.85$ (blue); power =$0.75$ (green); power =$0.5$ (cyan); power =$0.25$ (magenta).}
\label{fig:preproc}
\end{figure}

One of the main challenges for generating scientific data through techniques developed for computer vision lies in the inherent difference between the dynamic ranges of the pixel intensities. The pixel intensities in a typical RGB image have a range from $0$ to $255$ while the energy deposited in detector cells covers more than $10$ orders of magnitude. We have implemented a pre-processing procedure aimed at reducing this dynamic range. Initial tests conducted taking the logarithms of the pixel intensities, resulted in the generation of highly distorted images. Taking a less drastic approach we calculate the power function of pixels intensities using an exponent smaller than one. A smaller exponent results in faster convergence but greater distortion in generated images, while a larger exponent slows down convergence yet retaining image quality. Figure~\ref{fig:preproc} shows how the value of the exponent ($p$) affected the distribution of the generated pixel intensities for the individual cells. The value of $p$ is adjusted to an optimum value of $0.85$, where a faster convergence is achieved while retaining an acceptable level of accuracy at both ends of the spectrum. The generated images are then post-processed by simply taking the inverse of the power function. The data is also subjected to a threshold of $0.2$ GeV for the total energy deposited in the event. This rejection is aimed at removing some spurious events with little or no energy deposition.

\subsection{Architecture}
\label{sec:arch}

\begin{figure*}[ht]
    \centering
    \includegraphics[scale=0.57, trim={0cm 7cm 4cm 0}, clip=true]{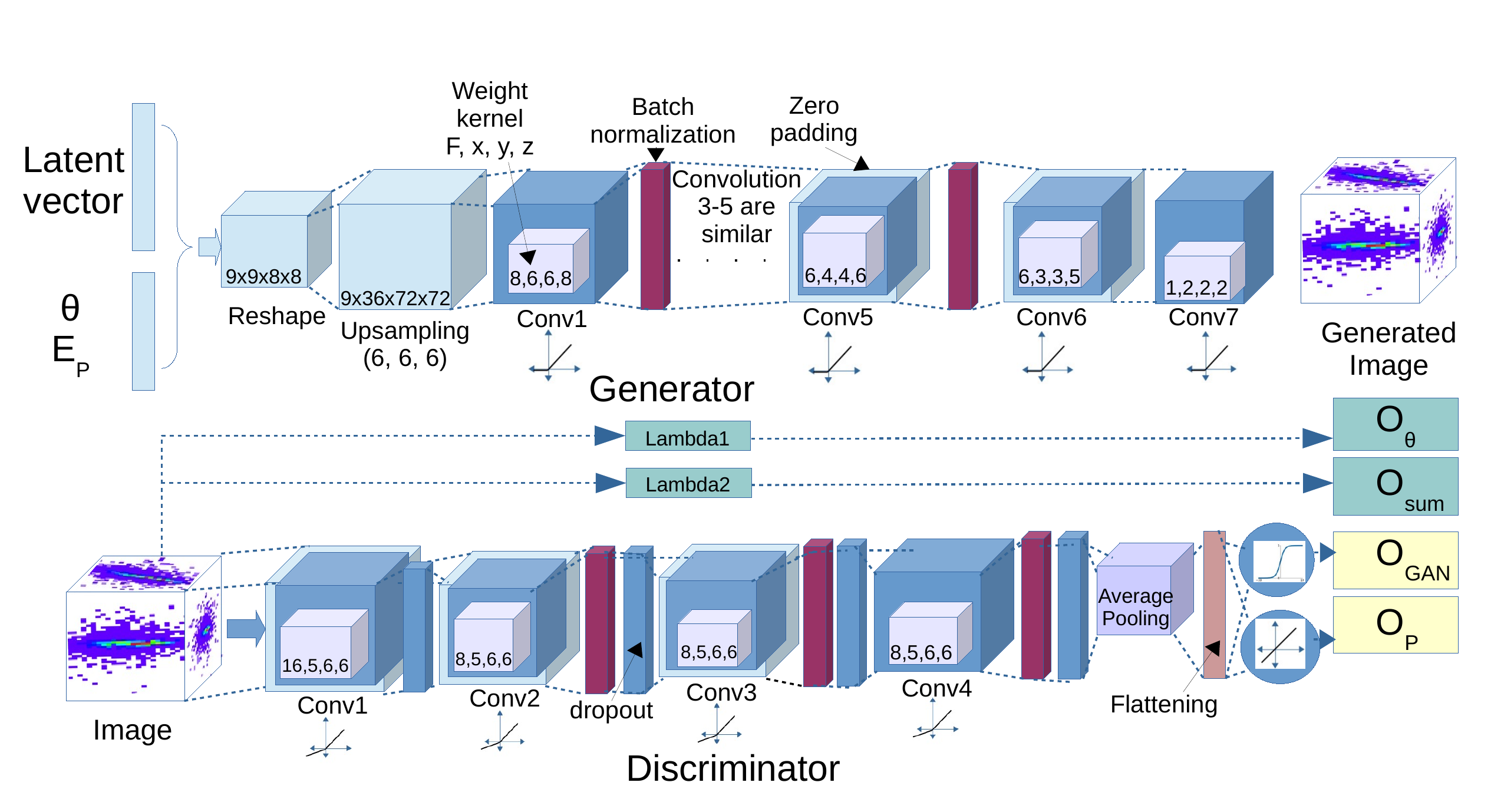}
    \caption{The 3DGAN architecture, see the text for details.}
    \label{fig:GAN_arch}
\end{figure*}

The 3DGAN architecture is presented in Figure~\ref{fig:GAN_arch}. The generator network implements stochasticity through a latent vector of $254$ random numbers drawn from a Gaussian distribution. The generator input includes $E_{P}$ and $\theta$ concatenated to the latent vector. The generator network then maps the input to a layer of linear neurons followed by seven 3D convolutional layers. The discriminator input is an image while the network has only four 3D convolutional layers. Batch normalization~\cite{BN} is performed after all except the first convolutional layer in the discriminator and the last two layers in the generator. The leakyRelu~\cite{leakyrelu} activation function is used for the discriminator hidden layers while the Relu~\cite{relu} activation function is used for the generator layers to induce sparsity. The discriminator uses dropout~\cite{dropout} of $20\%$ for regularization and a single average pooling layer after the last convolutional layer since additional pooling layers result in substantial loss of performance. 

The discriminator network has two trainable outputs: a sigmoid neuron predicts the $O_{GAN}$ and a linear neuron $O_{P}$ predicts $E_{P}$. The other two additional outputs are simple analytical measurements: $O_{sum}$ is the total deposited energy and $O_{\theta}$ is the measured incident angle (geometrical angle of the shower energy depositions). These non-trainable outputs represent physics-based constraints. 

\subsection{Loss Function}
\label{sec:loss}

The 3DGAN loss function is the weighted sum of individual losses pertaining to the discriminator outputs and constraints. The domain-related constraints are essential to achieve a high level of agreement over the very large dynamic range of the image pixel intensity distribution. Equation~\ref{eq:loss} presents the discriminator loss related to the output $O_{GAN}$ as $L_G$, the loss related to the output $O_{sum}$ as $L_{E}$, the output $O_{\theta}$ as $L_{A}$ and the predicted $E_{P}$ as $L_P$ balanced by corresponding weights $W$. 

\begin{equation}
   L_{3DGAN}  = W_{G}L_{G} + W_{P}L_{P} + W_{A}L_{A} + W_{E}L_{E}
\end{equation}
\label{eq:loss}

The $L_P$ and $L_{\theta}$ both provide feedback on how well the generated images correspond to the input conditions. The loss $L_{E}$ ensures energy conservation.  $L_{G}$ is evaluated as binary cross-entropy. $L_{P}$ and $L_{E}$ are implemented on mean percentage errors, while $L_{A}$ as mean absolute error. The generator loss is implemented as the inverse of $L_{G}$ together with the auxiliary losses and constraints. The weights (presented in Appendix~\ref{app:params}) are considered as hyperparameters and chosen to balance the loss ranges and their relative importance (in this case the loss $L_G$ is given higher priority as compared to the auxiliary losses).

\subsection{Training}

The 3DGAN training is inspired by the concept of transfer learning. The GAN could not converge for the highly complex multivariate distribution directly thus a two-step training is applied. In order to successfully train the network, we reduce the complexity by training the GAN first for electron events having $E_P\sim~U(100, 200)$~GeV. After the GAN converges, the same trained model is further trained with the data from the whole $E_P$ range of $E_P\sim~U(2, 500)$~GeV. The first training step exploits $137,342$ electron events. The GAN is then trained for the larger $E_P$ range, utilizing a much larger size of training data ($400$ k  events) from each particle type (electrons, photons, and neutral pions). The train and test losses are evaluated on the data divided in a ratio of nine to one. The first training step is run for $130$ epochs (2 hours per epoch on GTX 1080) while the second step is run for $30$ epochs (4 hours per epoch on GTX 1080). Finally, the best network is selected according to the minimum relative error for the $SF$ on additional validation data ($20k$ events are filtered around specific $E_P$ bins). This last step is aimed to further improve the accuracy for the $SF$. 

%
%

\begin{algorithm}

\SetAlgoLined
 initialize\\
 \For {number of epochs}{
  \#Training\\
  \For{all batches in the training data}{
  get real $E_P$, $\theta$, $E_{sum}$ and image batches from data\\
  latent batch $\sim \mathcal{N} (0 \,, 1)$\\
  generator input = $E_P \concat \theta \concat $ latent\\
  generate fake events for the same $E_P$ and $\theta$\\
  train discriminator on real batch\\
  train discriminator on fake batch\\
  \For{2 times}{
   latent batch $\sim \mathcal{N} (0 \,, 1)$\\
   generator input = $E_P \concat \theta \concat$ latent\\
   use real $E_P$ and $\theta$ for fake events\\
   train the generator (minimizing the discriminator loss on the generated images)\\
  }
  
   $\#$ Testing\\
   Evaluate discriminator on real and fake data\\
   Evaluate generator using the inverse of discriminator loss on generated data
  }
 }
 \caption{Training the 3DGAN model}
 \label{alg:var_train}
\end{algorithm}

\begin{figure}
\centering
\includegraphics[width=0.6\textwidth, trim={0cm 8cm 7cm 0cm}, clip=true]{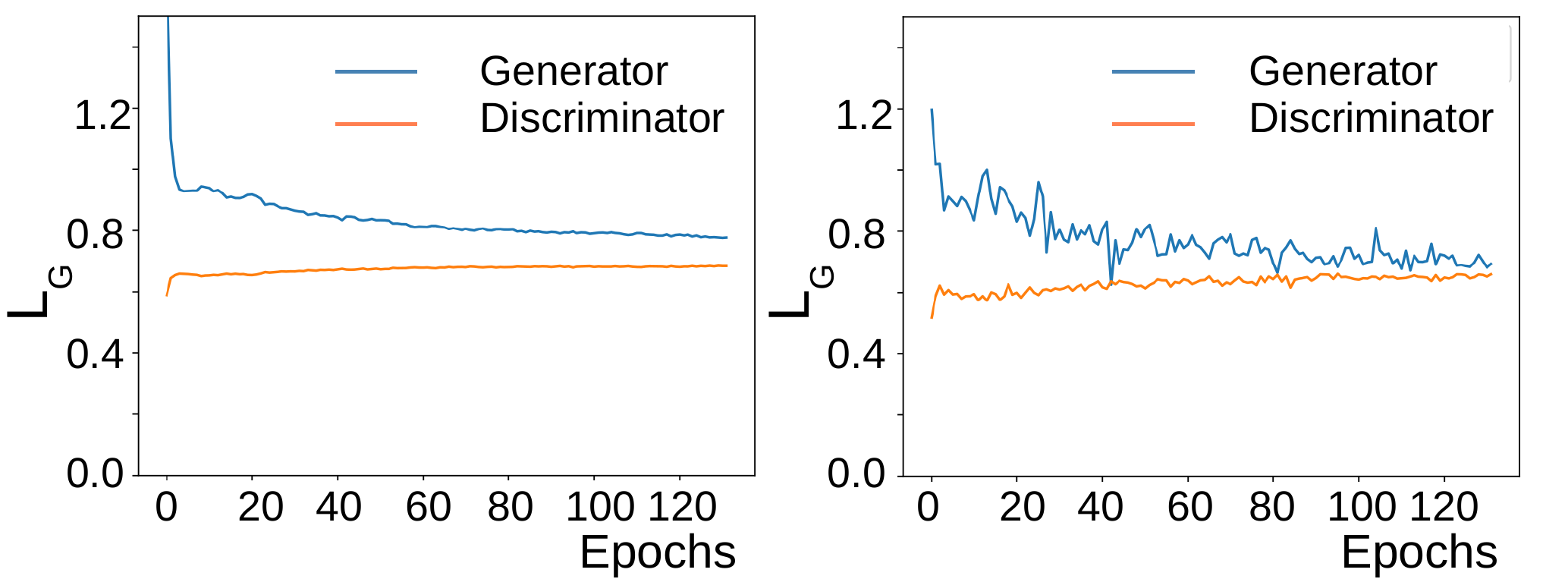}
\includegraphics[width=0.302\textwidth, trim={0cm 7.5cm 17.25cm 0.58cm}, clip=true]{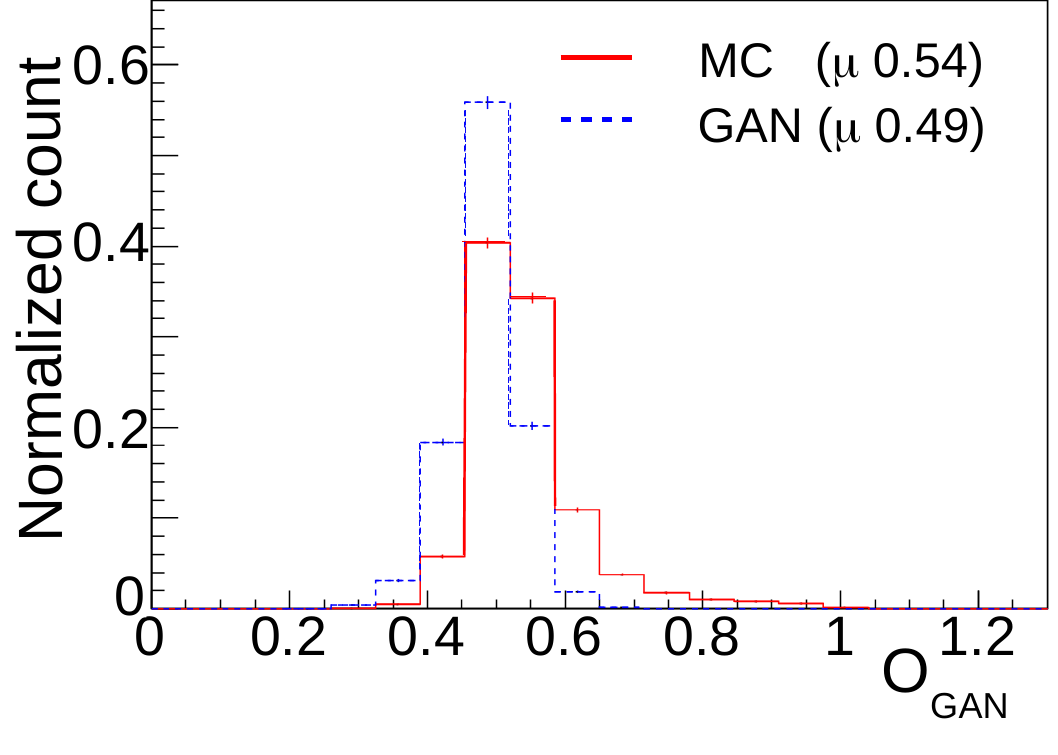}
\caption{The $L_{G}$ losses for the generator (blue) and the discriminator (orange) converge for the first training step with restricted $E_{P}$ range. (Left) training losses. (Middle) test losses. (Right) the discriminator output $O_{GAN}$ for MC (red) and GAN (blue) images is similar and the discriminator is confused.}
\label{fig:training}
\end{figure}

The training process for each epoch is presented by the Algorithm~\ref{alg:var_train}. For each training iteration the discriminator is trained twice: once on a batch of real data, and next on a batch of generated data. For a balanced approach, the generator is also trained twice while freezing the weights of the discriminator. The RMSProp~\cite{rmsProp} optimizer is utilized to train the network through Stochastic Gradient Descent. Figure~\ref{fig:training} shows the $L_{G}$ losses associated to the discriminator (blue) and the generator (orange). It can be seen that the loss for the discriminator increases, while the loss for the generator increases till both losses are converged at around $0.6$ (log(4)). At this point, the $O_{GAN}$ output for both the data (red) and the GAN images (blue), shown in Figure~\ref{fig:training} right panel has similar distributions centered around $0.5$. The discriminator is indeed confused and the GAN converges.  

\subsection{Generation Time}

The 3DGAN greatly reduces the simulation time. Table~\ref{tab:timing} compares the time taken to generate a shower using Monte Carlo and GAN. The inference time is around $4$ ms/particle on GeForce GTX 1080. A similar shower can be simulated using the GEANT4 in about $17$ second per particle on an Intel Xeon 8180 (currently it is not possible to run a full Geant4-based simulation on GPUs). The speedup of many orders of magnitude is achieved.

\begin{table}[]
    \centering
    \caption{Inference Timing for 3DGAN}
    \begin{tabular}{p{0.2\textwidth}  p{0.4\textwidth} p{0.15\textwidth} p{0.1\textwidth}}
         Method & Platform & Time/shower (msec) & Speedup \\
         \hline
         \hline
         Monte Carlo& \multirow{2}{*}{\parbox{0.4\textwidth}{2S~Intel~Xeon~Platinum~8180}} & $17000$ & $1.0$\\
         3DGAN CPU &  & $16$ & $2500$\\
         3DGAN GPU &  GTX 1080 & $4$ & $4250$\\
         
         \hline
         \hline
     \end{tabular}
    
    \label{tab:timing}
\end{table}{}

\section{Results and Discussion}
\label{sec:results}

The performance assessment for GAN models is a subject of much debate and diverse viewpoints~\cite{empirical}. The GAN evaluation is nontrivial due to the intractable probability densities and thus is mainly sample-based and application-specific. We have validated the realism and diversity of our generated data from several independent viewpoints, such as the output of a third-party neural network and image quality assessment as presented previously~\cite{access2021} but the current work focuses mainly on the physics-based comparison to a Monte Carlo simulation.

The particle showers have specific characteristics due to the underlying physics processes, depending on the detector material and the type, energy, and direction of the particle initiating the shower. We validate these characteristics as a function of our inputs by dividing the data in $5$ GeV $E_P$ and $0.1$ radian ($5.73^\circ$) $\theta$ bins. To ensure an unbiased comparison, GAN events are generated with the same $E_{P}$ and $\theta$ values as the GEANT4 events. The bin-wise comparison of each physics-based feature, results in hundreds of histograms, for each particle type. We present here a selected subset of the detailed and exhaustive validation that we consider to be the most essential and representative of performance. The results presented in this section are also alternated among different bins and particle types, in order to convey the overall level of accuracy.

\subsection{Visual Inspection}

\begin{figure}
\centering
\includegraphics[width=0.95\textwidth, trim={0cm 0cm 0cm 1cm}, clip=true]{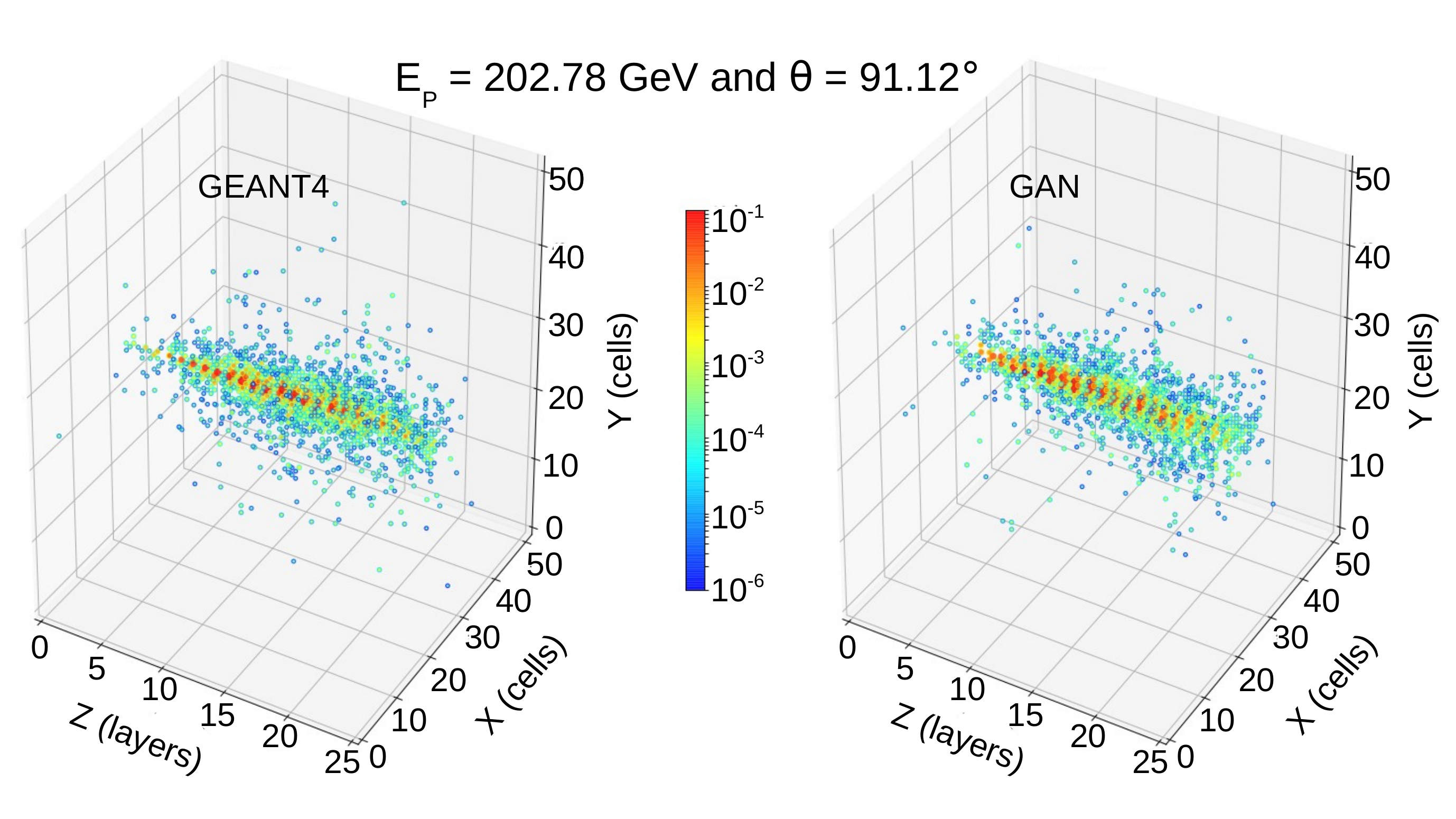}
\caption{GEANT4 vs. GAN electrons showers with $E_p = 202.78$ GeV and $\theta = 91.12^\circ$.}
\label{fig:proj_3d}
\end{figure}

\begin{figure}
\centering
\includegraphics[width=0.95\textwidth, trim={0cm 4.5cm 2cm 0cm}, clip=true]{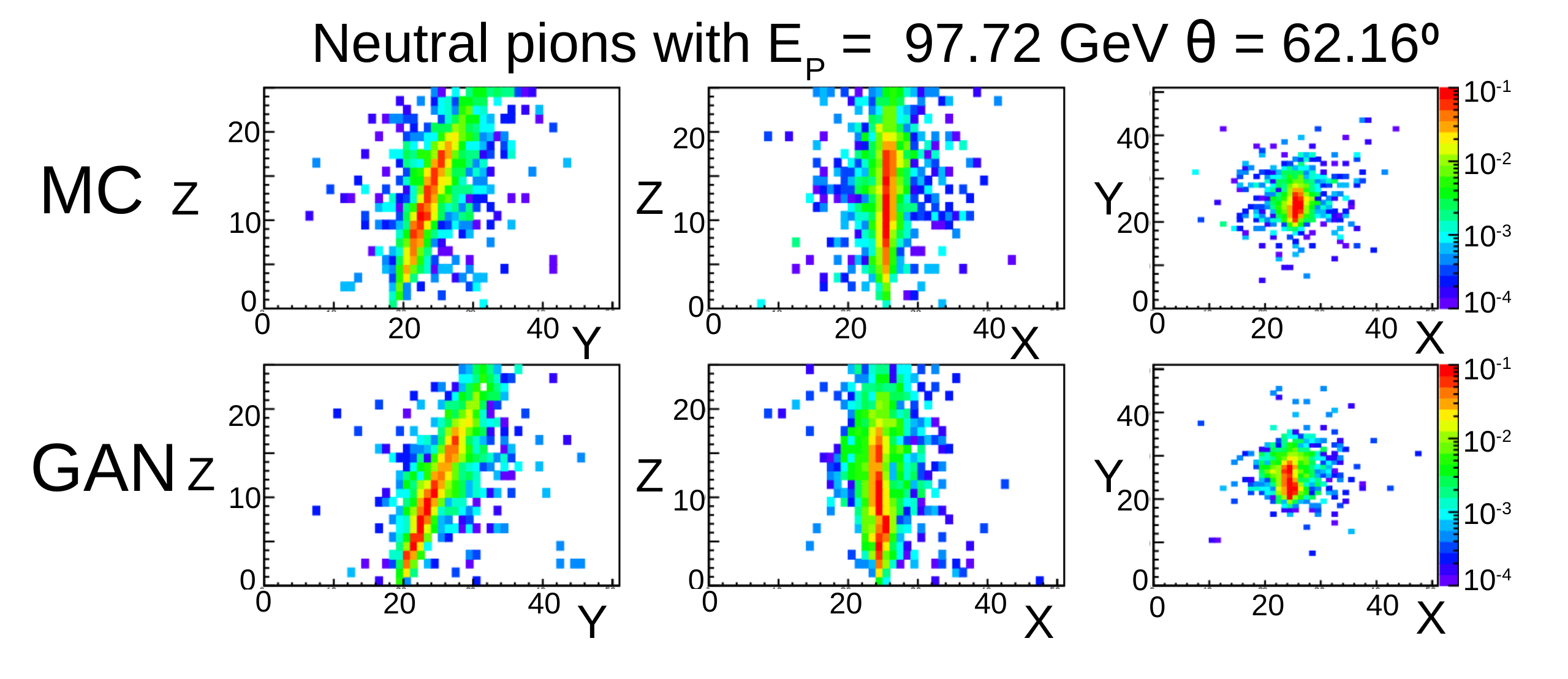}
\includegraphics[width=0.95\textwidth, trim={0cm 4.5cm 2cm 0cm}, clip=true]{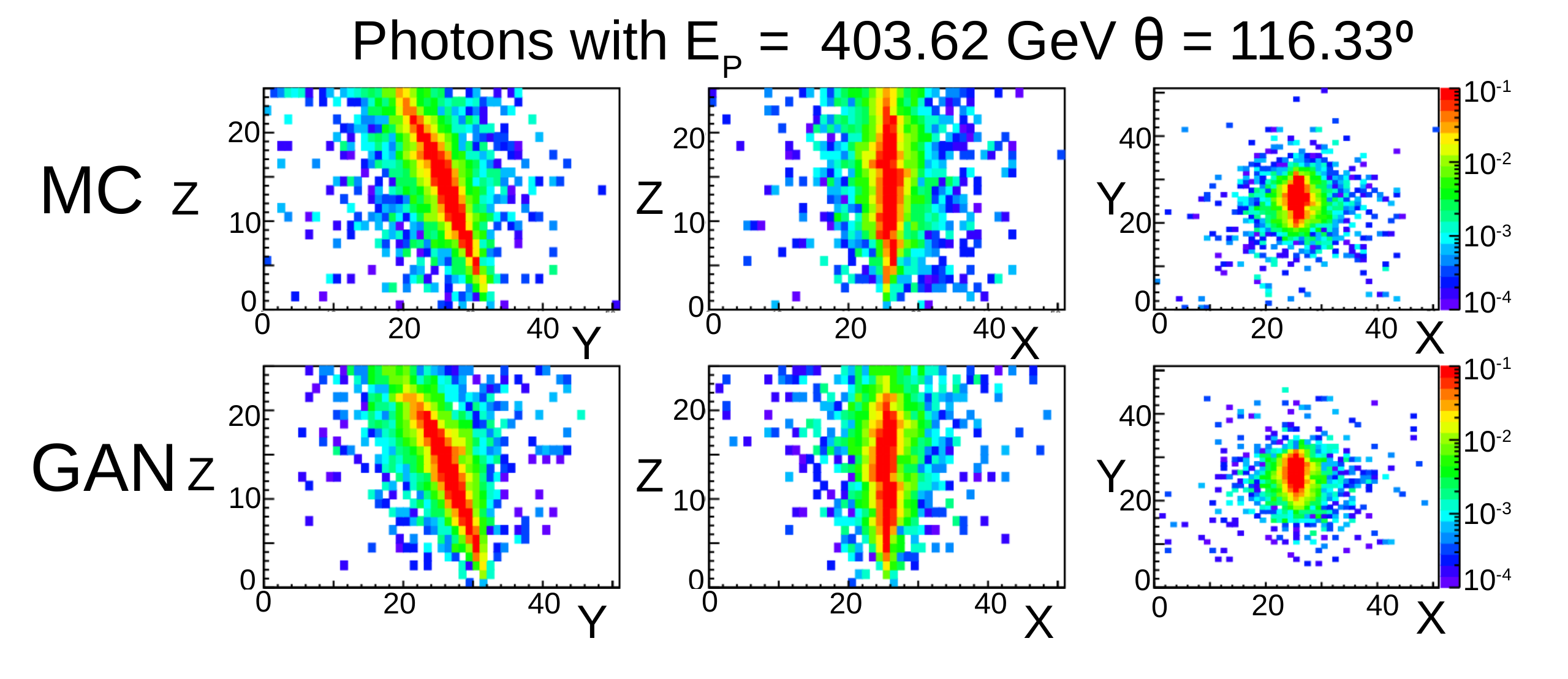}
\caption{Projections of Monte Carlo vs. GAN events on the YZ, XZ and XY planes. Top)  neutral pions with $E_P= 97.72$ GeV and $\theta = 62.16^\circ$. Bottom) photons with $E_P= 403.62$ GeV and $\theta = 116.33^\circ$.}
\label{fig:proj_var}
\end{figure}

An initial qualitative assessment can be performed by comparing the events visually. Figure~\ref{fig:proj_3d} shows an example of a 3D electron event. The event on the left has been generated by GEANT4, while the event on the right has been generated by the GAN for the same input values. It can be observed that both events have very similar visual characteristics while retaining uniqueness for individual cell deposits. The graphical projections on different planes further illustrates the shower correlation to the incident angle ($\theta$) and energy ($E_{P}$). Figure~\ref{fig:proj_var} compares the projections of the GEANT4 showers to the corresponding GAN showers. The top panel presents neutral pion events, while the bottom panel display photons events, with the $E_P$ and $\theta$ from both ends of the spectrum. The GAN images appear similar to the respective GEANT4 images with the deposited cell energies correlated to the input  conditions while retaining stochasticity, for all particle types and input conditions.

\subsection{Particle Shower Features}
\begin{figure}
\centering
\includegraphics[width=0.95\textwidth, trim={0cm 1cm 0cm 0}, clip=true]{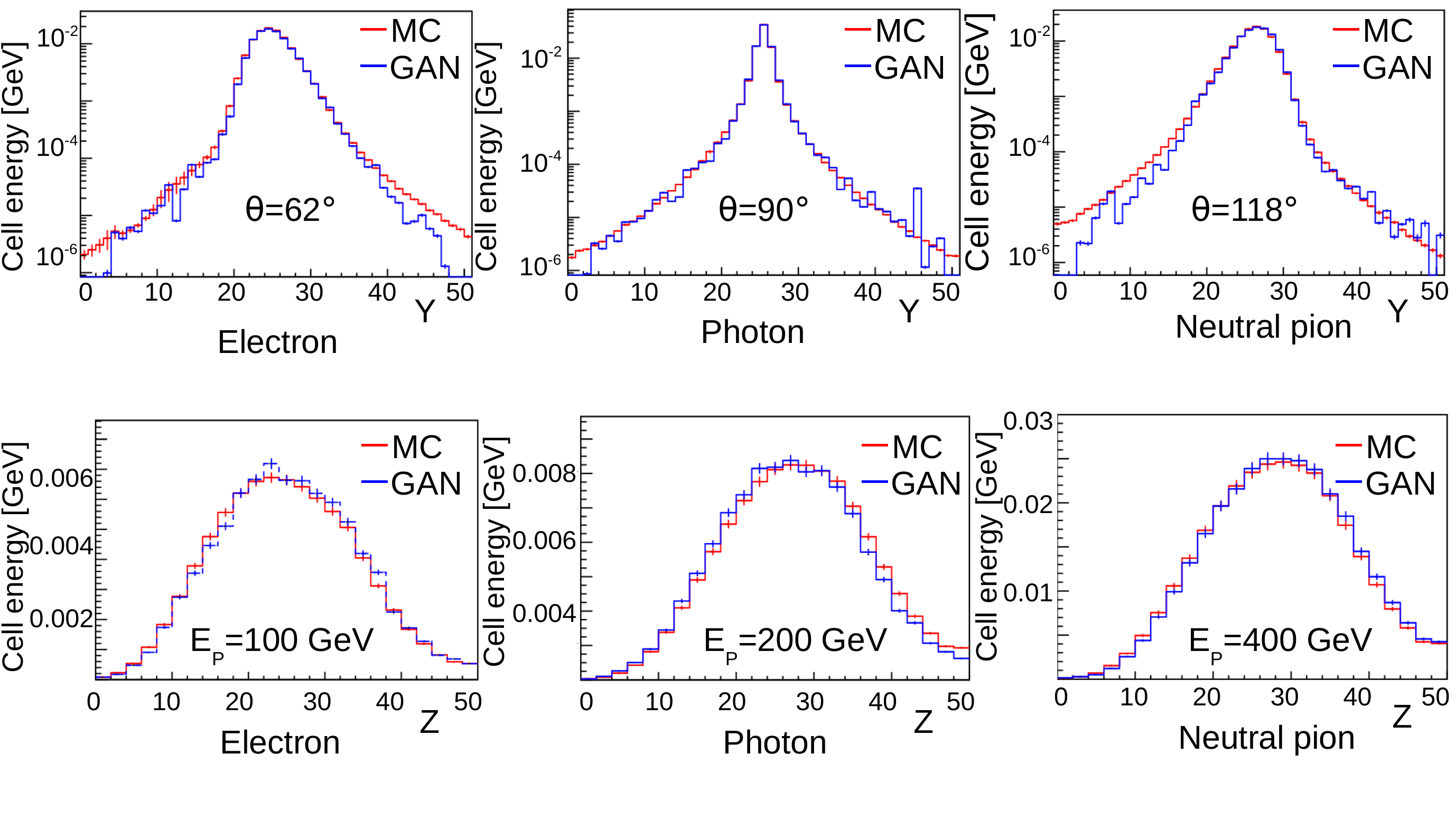}
\caption{Shower Shapes for the GEANT4 vs. GAN events as a function of inputs. Top row) longitudinal shower shapes along the $Y$ axis: left) electrons with $\theta$ in the $62^{\circ}$ bin; mid) photons with $\theta$ in the $90^{\circ}$ bin; right) neutral pions with $\theta$ in the $118^{\circ}$ bin. Bottom row) transverse shower shapes: left) electrons with $E_{P}$ in the $100$ GeV bin; mid) photons with $E_{P}$ in the $200$ GeV bin; right) neutral pions with $E_{P}$ with the $400$ GeV bin.}
\label{fig:GAN_shapes}
\end{figure}

Shower shapes define the structure of the deposited energy distribution, as a shower develops through the detector material. The profiles of the energy deposition along the detector axes are important observables related to the shower geometry and crucial for most particle identification techniques. We would like to point out that these geometrical features are not included in the 3DGAN loss function as presented in Equation~\ref{eq:loss} and are learned by the GAN implicitly. Figure~\ref{fig:GAN_shapes} presents the shower shapes for the $X$, $Y$ and $Z$ axes as a function of $\theta$ and $E_{P}$. In order to summarize the performance for all particle types, we present a different particle in each column: the shapes for electrons in the first column from the left, the photons in the middle column, and the neutral pions in the rightmost column. The top row presents the transverse shape distribution for the $Y$ axis corresponding to the different $\theta$ bins since the $Y$ axis profile is most relevant for $\theta$. The plots are displayed in the log scale to enhance the sparse distributions along the tails. The second row presents the shapes along the $Z$ axis (longitudinal direction) in linear scale for the different $E_{P}$ bins. The network is able to reproduce a similar shape distribution as the GEANT4 showers, furthermore, the network can correctly relate it to the inputs. In the transverse profiles, some discrepancies are observed in the log scale. These discrepancies occur at the volume edges, where smaller energy depositions occur. This region is also highly sparse and outside the main body of the shower with expected energies well below $0.1$ MeV, which is comparable to the pedestal values. 

\begin{figure}
\centering
\includegraphics[width=0.95\textwidth, trim={0cm 9cm 0cm 0cm}, clip=true]{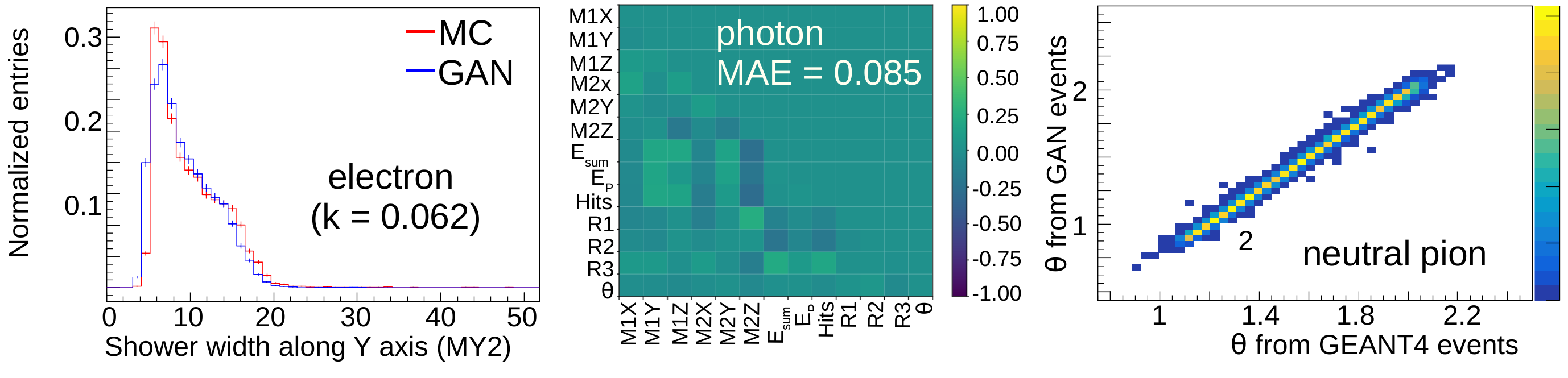}
\caption{The GAN vs. GEANT4 shower features. Right) the shower width (M2) along the $Y$ axis for electrons. Mid) difference between internal correlations present between physics features and the inputs for photons. Left) the correlation between the measured angle from GEANT4 events and the GAN events generated for the same $\theta$ values for the neutral pions.}
\label{fig:feat_1}
\end{figure}

Moments are another aspect of the shower geometry. The GEANT4 showers are all centered on the barycenter of the energy deposition by design thus the first moment ($M1$) defining the shower center is easily replicated by GAN. Therefore, we present here the performance related to the second moment ($M2$) having a more complex distribution depending on both  $E_{P}$ and $\theta$. Figure~\ref{fig:feat_1} left panel presents the distribution of the second moment or the width of the shower for electrons. Here it can be appreciated that the GAN has learned the non-Gaussian width distribution. The mid plot shows the difference between internal correlation present between the shower inputs ($E_{P}$ and $\theta$) and the shower features (shapes, moments, total deposition, hits, and ratios of energy deposition in different parts of the shower) for GEANT4 and GAN photons. The GAN showers are able to reproduce the internal correlations present between the different shower observables. The right panel displays the close agreement between the $\theta$ measured from GEANT4 neutral pion events and that measured from the GAN events generated for the same $\theta$ values. 

\begin{figure*}
\centering
\includegraphics[width=0.95\textwidth, trim={0 9cm 0cm 0cm}, clip=true]{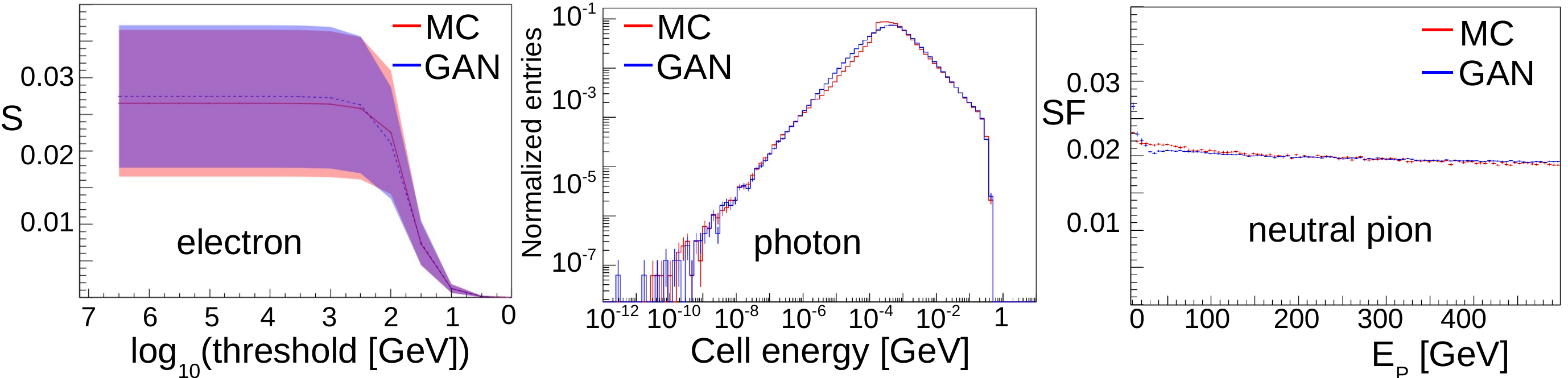}
\caption{Shower features related to the pixel intensities (cell depositions) for the GEANT4 (red) vs. GAN (blue) events. Left) sparsity level as a function of cutoff threshold for electrons. Mid) distribution of cell energy deposits for photons. Right) sampling fraction for neutral pion showers.  }
\label{fig:feat2}
\end{figure*}

\begin{figure}
\centering
\includegraphics[width=0.95\textwidth, trim={0cm 3.8cm 0cm 0cm}, clip=true]{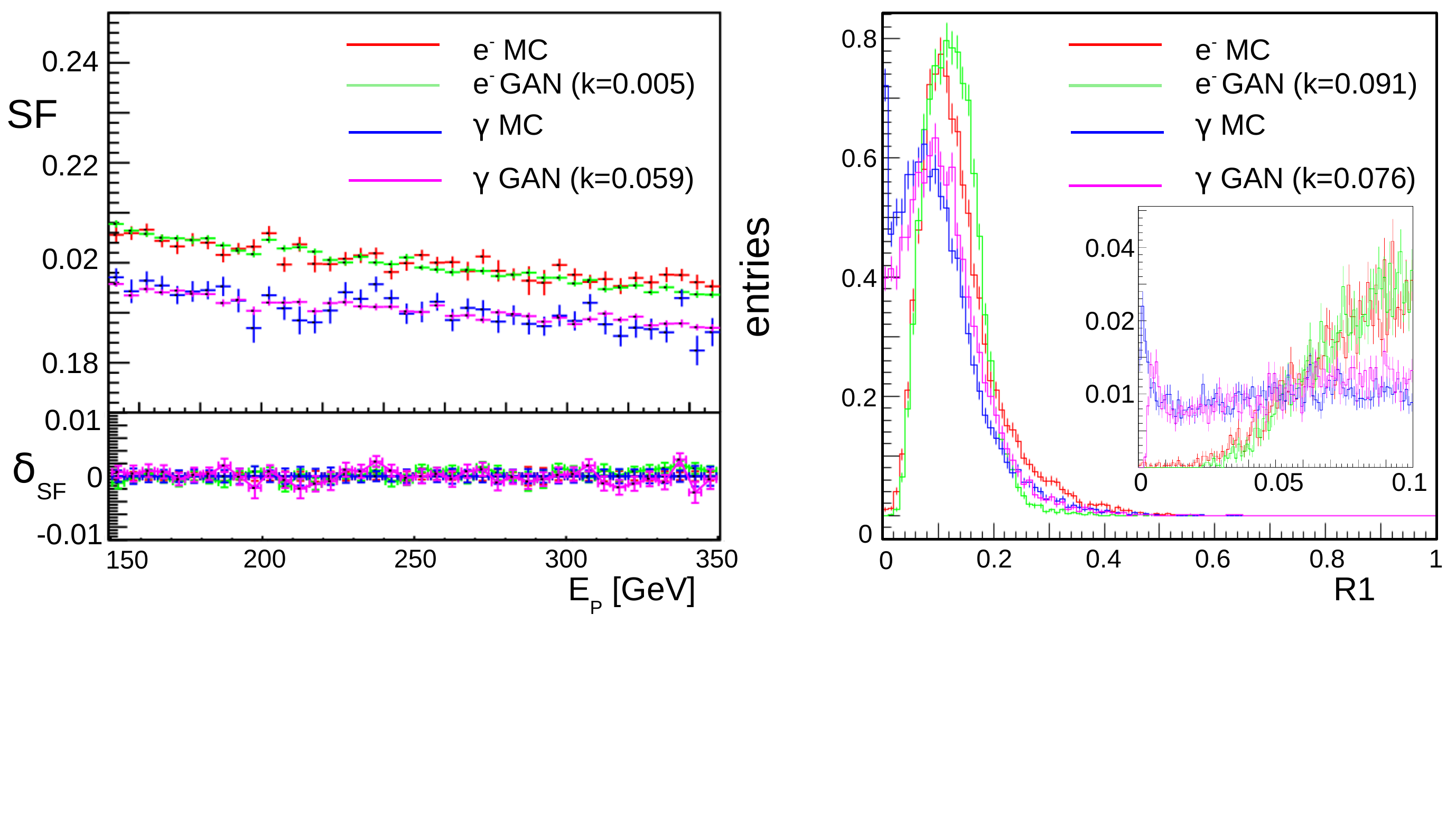}
\caption{The GAN could successfully learn the photon features after further training a network generating electrons with reduced $E_P$ range. Left) the $SF$ for photons is lower than electrons both in GEANT4 events and GAN events. The bottom panel shows the relative error ($\delta_{SF}$) for the sampling fraction. Right) the distribution of the energy fraction deposited in the first part of the shower ($R1$) for the electron and photon showers. The distribution for $R1 < 0.1$ (zoomed in the inset) has very different distributions for electrons and photons. The GAN also demonstrates a similar behavior.}
\label{fig:feat_diff}
\end{figure}

The energies deposited in detector cells are the pixel intensities of our images. The images are mostly empty, centered around a shower. The energies are deposited only in around $20\%$ of the cells. Figure~\ref{fig:feat2} left panel shows sparsity (S) as the fraction of cells with some deposition against the threshold used for cutoff. The GAN images have similar sparsity distributions as the GEANT4 events, without specifically constraining the image. The distributions for cell energy depositions (photon) shown in the mid panel have a similar shape for GEANT4 and GAN events. We had reported that a sharp, vertical drop around $0.2$~MeV was present in the GEANT4 cell energy distribution that the GAN could not learn, yet tried to smooth out in the best manner~\cite{3dgan_icmla}. Since then a recent work~\cite{high2020} also recognizes this feature and tries to replicate the effect by additional post-processing on the generated images. They note that their network improves the performance on the simulation of the correct pixel distribution at the cost of reduced performance for other features. We also report a reduction in performance when constraining the pixel intensity distribution through our loss function. We believe in the future the concept of ensembling~\cite{ensambling} can be used to employ two networks to generate the pixels above and below this region. The sampling fraction ($SF$) is an important characteristic of the detector response. Figure~\ref{fig:feat2} left panel presents $SF$ for neutral pions. The $SF$ is presented as a function of $E_{P}$. There is a close agreement for most of the input range with some discrepancies at low energies, where events are highly sparse with low cell energy deposits. 

The characteristic features for the electromagnetic showers are faithfully reproduced in the GAN-generated showers for different particle types. We test if the generated showers for different particle types are mutually distinguishable through their corresponding features. This is crucial as the final networks for all particle types, use the same initial weights trained to generate electron-induced showers for a reduced energy range. The photon-initiated showers have some minor differences from electron-initiated showers. The photons penetrate more distance into the detector material before starting to interact ~\cite{Fabjan2020, calo_presentation}. This effect can be evaluated by studying the $SF$ and the fraction of energy deposited in the first ($8$ cells along the $Z$ axis) part of the shower. Figure~\ref{fig:feat_diff} left panel shows the profiles of $SF$ as a function of $E_P$ for the central region of the spectrum. The $SF$ for photons is lower than electrons for a similar value of $E_P$. The right panel compares the distribution of energy deposited in the first part of the shower for electrons and photons. The photons present more entries for the region where the fraction of energy deposited in the first part of the shower $R1$ is less than $10\%$. The GAN-generated photons clearly demonstrate these identifiable features.

\section{Simulating Charged Pions}
\label{sec:chpions}

The charged pions deposit a much smaller part of their energies in the ECAL while most of the energy is deposited in the HCAL. The current project is only limited to the ECAL data due to the limitation of the computing resources, thus the work for charged pion is only a preliminary study. A more accurate approach will also need to incorporate data from HCAL. We will present the results of our study to lay the foundation for any future work.



\begin{figure}
\centering
\begin{tabular}{c c}
    \includegraphics[width=0.95\textwidth, trim={0cm 9.5cm 0cm 0}, clip=true]{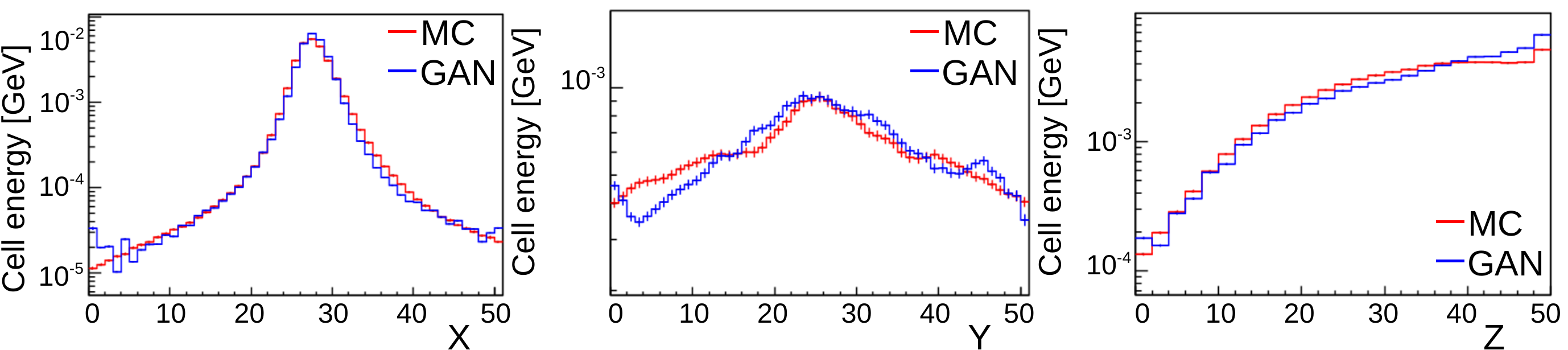}
\end{tabular}
\caption{Shower Shapes for GEANT4 vs. GAN charged pion events along $X$ (left), $Y$ (mid) and $Z$ (right) axes with random $E_P$ and $\theta$.}
\label{fig:shapes_chpi}
\end{figure} 

\begin{figure}
\centering
\includegraphics[width=0.95\textwidth, trim={0cm 14cm 0cm 0cm}, clip=true]{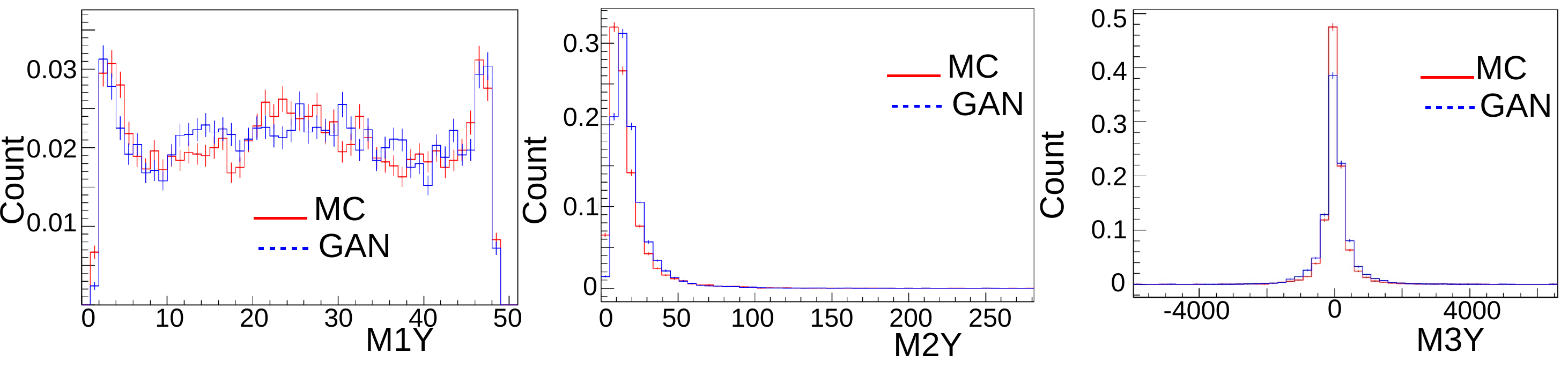}
\caption{Shower moments along $Y$ axis for GEANT4 vs. GAN charged pions. Left) $M1Y$ (shower center). Mid) $M2Y$ (shower width). Right) $M3Y$ (shower skewness). }
\label{fig:moments_chpi}
\end{figure}

The transfer learning approach could not be extended to charged pions as the showers have a very different distributions. The GAN is trained for the full $E_P$ range of the charged pions, directly from random weights, for about $200$ epochs using ($300k$) events. There is a great diversity in the charged pion events, and most of the events have low energy deposition in the ECAL. Since visual inspection will not be helpful to understand the performance, thus we present only the distributions of physics-based features. Figure~\ref{fig:shapes_chpi} presents the overall shower shapes in log scale. The shape distributions along the $X$ and $Z$ axis show slightly better performance as compared to the $Y$ axis, probably due to the higher variance present in this dimension. Figure~\ref{fig:moments_chpi} compares the distribution of the first three moments defining the shower center, width, and skewness along the $Y$ axis, display similar distributions. 


\begin{figure}
\centering
\includegraphics[width=0.98\textwidth, trim={0.5cm 9cm 0.5cm 0cm}, clip=true]{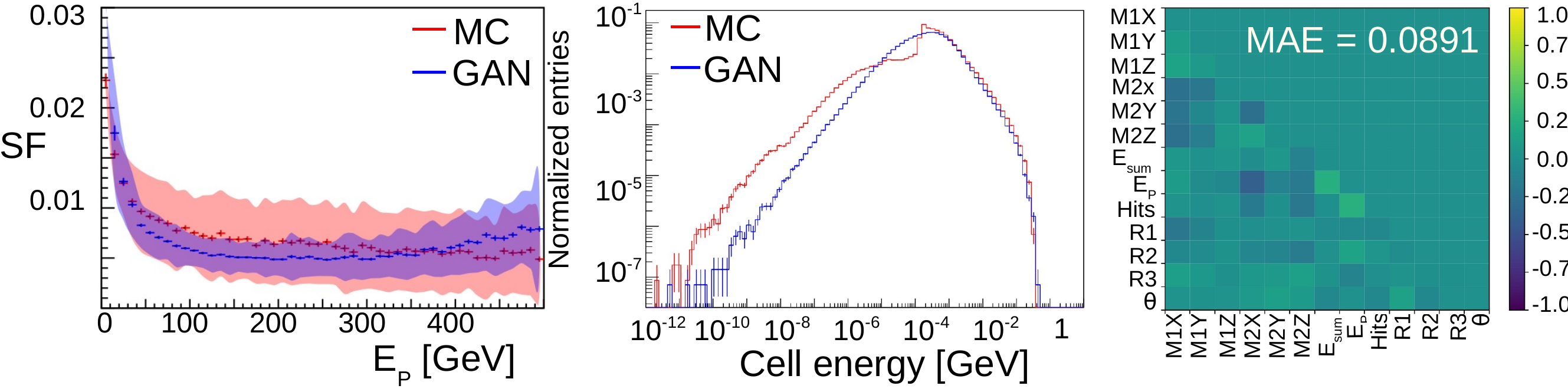}
\caption{Shower features for GAN vs. GEANT4 charged pion events. Left) $SF$ as a function of $E_P$ with shaded area representing standard deviation. Mid) pixel intensities with log yscale. Right) difference between internal correlations among shower features.}
\label{fig:feat_chpi}
\end{figure}

\begin{figure}
\centering
\includegraphics[width=0.85\textwidth, trim={0cm 0cm 6.1cm 0cm}, clip=true]{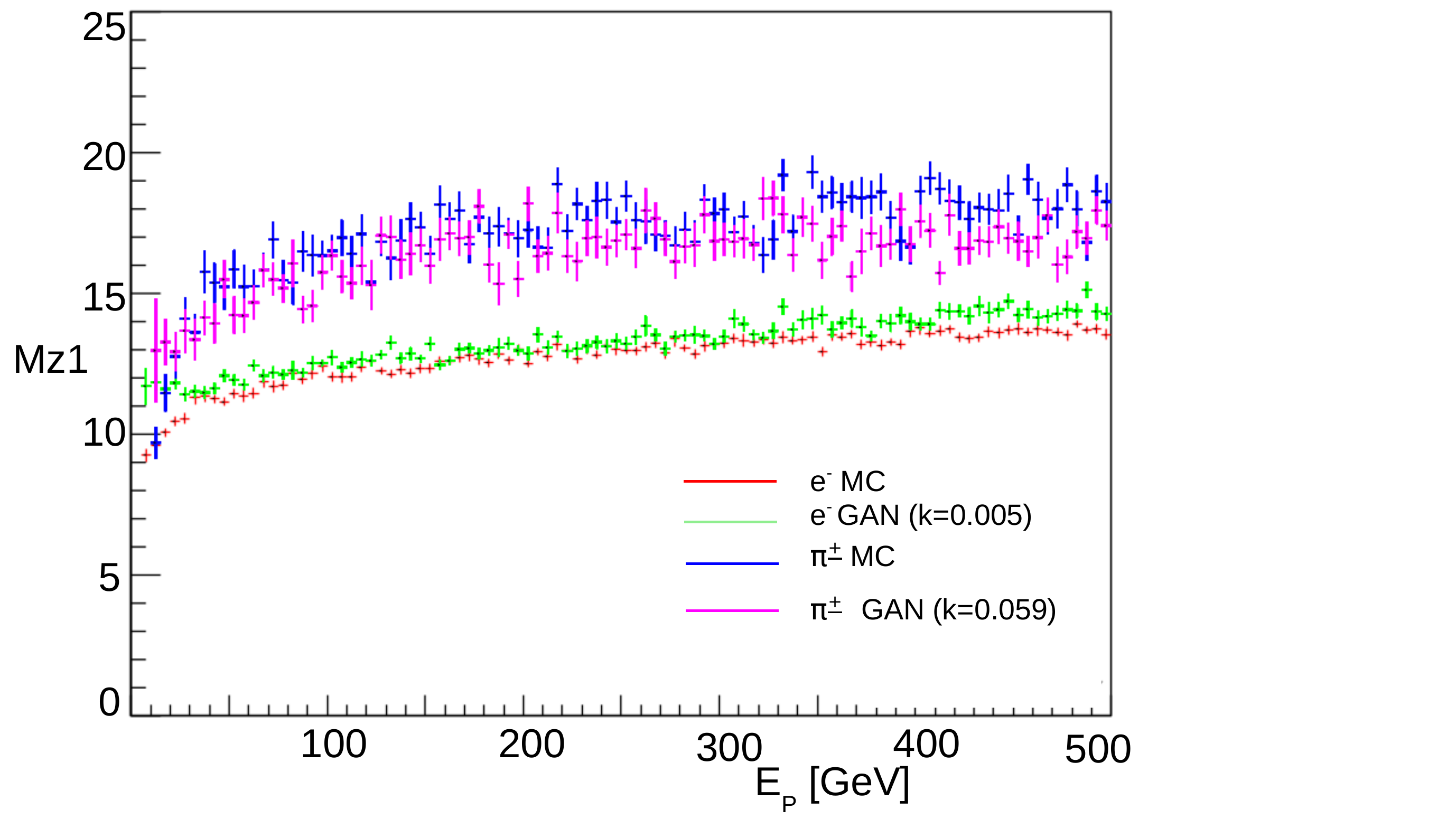}
\caption{The distribution for the first moment along the $Z$ axis (Mz1) for electrons and charged pions. The barycenter of deposition along the $Z$ axis shows very different distributions for both particles with the charged pion showers starting much later.}
\label{fig:momentz}
\end{figure}

Figure~\ref{fig:feat_chpi} left panel displays the sampling fraction for GEANT4 and GAN charged pion events. It can be seen that there is a difference between the means of the two distributions, while the shaded area representing the standard deviation shows some overlap. The mid panel presents the distribution of the pixel intensities that are energies deposited in the calorimeter cells. There is a deterioration in performance as compared to other particles, particularly the effect of the presence of a cut around $0.2$ MeV (see Section~\ref{sec:results}) is more pronounced. The left panel presents the difference in internal correlations between physics-based features like shapes, moments, hits, $E_P$ and $\theta$ for the GEANT4 and GAN events. There is a less than $10\%$ error for the correlations, even given the highly diverse and incomplete showers. In order to convey an idea of the diversity present in these showers, we investigate the barycenters of the shower energy depositions along the longitudinal axis (Z) for electrons and charged pions, as a function of $E_P$. Figure~\ref{fig:momentz} compares the first moment for the GEANT4 and GAN events. The charged pion showers show greater diversity and depth as compared to electrons, also apparent in the GAN events.

\section{Rare modes}
\label{sec:rare}

\begin{figure}[]
\centering
    \includegraphics[width=0.85\textwidth, trim={0.5cm 0cm 0cm 0.1cm}, clip=true]{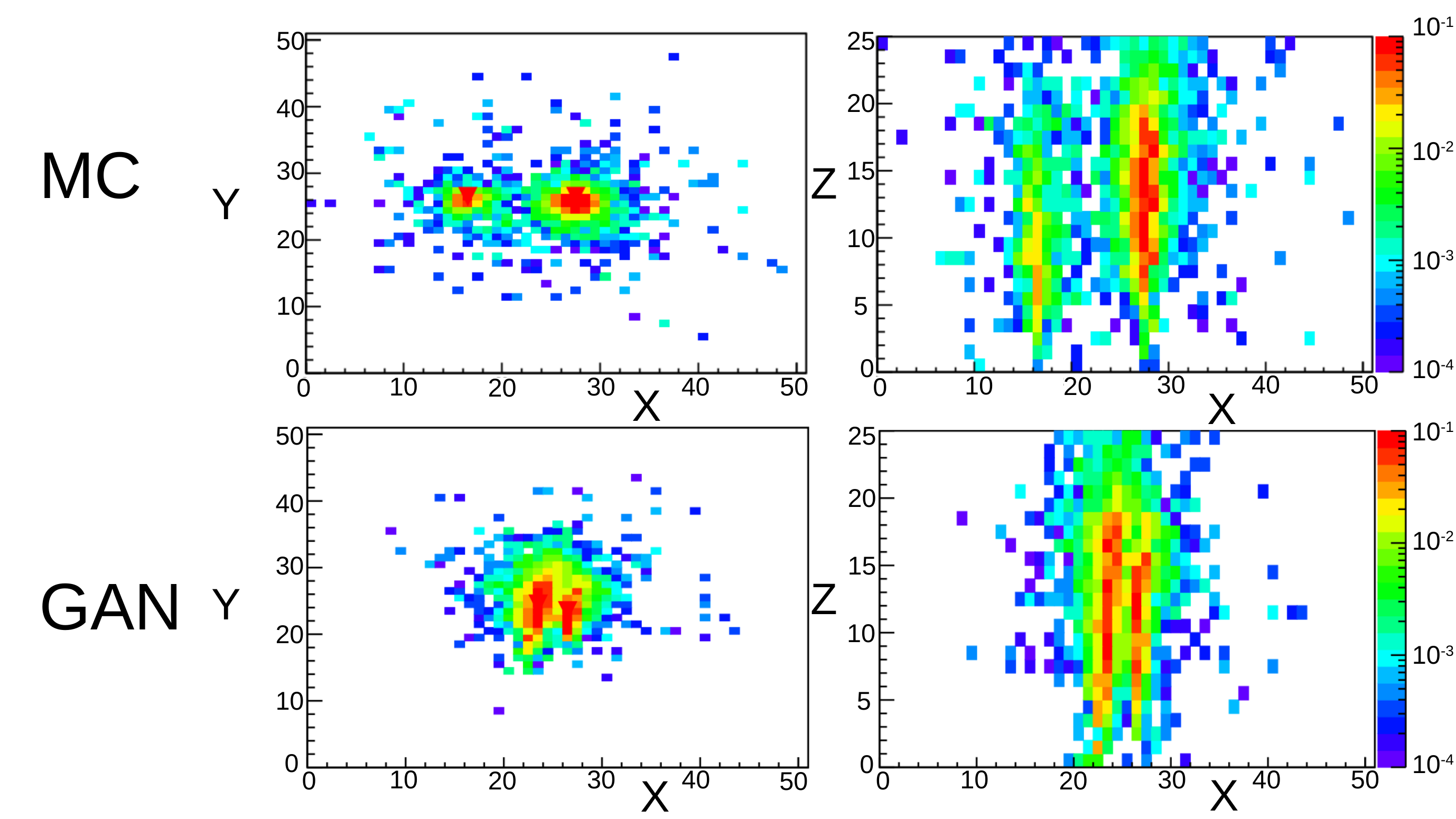}
\caption{Graphical projections for the $XY$ and $XZ$ planes for MC events with high discriminator probability of being real images; top) GEANT4 event; bottom) GAN event. }
\label{fig:outliers_g4}
\end{figure}

The discriminator assigns a higher probability of being realistic to images exhibiting features that the GAN cannot reproduce correctly . We visually investigate such events having $O_{GAN}$ value greater than $0.6$, with the help of graphical projections. It is observed that most of these events manifest rare modes in data like pre-showering, late showering, and incorrectly centered events. As other modes are found to be even rarer, thus only the early showering events are further investigated. Figure \ref{fig:outliers_g4} top row presents an example of a GEANT4 pre-showering event. The particles that start depositing their energies before entering the calorimeter volume have multiple particles striking the detector surface, thus resulting in multiple branches. Only a few percent of the Monte Carlo samples present such behavior and the percentage decreases with increasing $E_P$. 


Figure~\ref{fig:outliers_g4} bottom row presents an example of a pre-showering event generated by GAN. It must be mentioned here that in the first training step more such events are generated. The performance for these rare modes deteriorates with further training for the full $E_P$ range due to a decrease in the percentage of such events for higher energies. In order to further improve the performance, methods like the ensembling~\cite{ensambling} can be explored.

\section{Training with GAN data}
\label{sec:gan_training}

We present a practical use case for the GAN-generated events. Triforce~\cite{Belayneh2019} is a deep learning model developed by a third-party study for the identification of particle type and primary energy for particle showers from the calorimeter dataset used for 3DGAN. We had previously employed the pre-trained Triforce DNN model for classification and regression of the GAN events~\cite{3dgan_icmla} and proved that the type and the primary particle energy for the GAN-generated events was correctly predicted. 

\begin{figure}
    \centering
    \includegraphics[width=0.9\textwidth, trim={0cm 7cm 2cm 0cm}, clip]{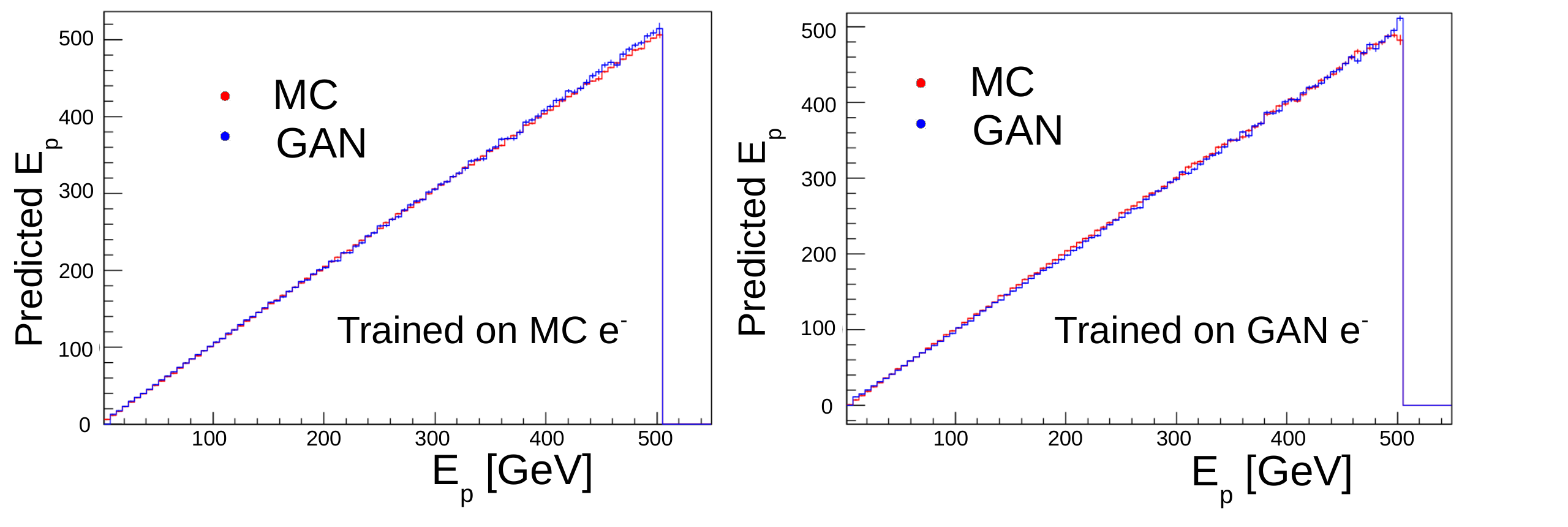}
    \caption{Performance of the network trained on GEANT4 electrons and charged pions (left) vs. that of the network trained GAN electrons and GEANT4 charged pions (right). The profile plot for true and predicted $E_P$ for the GAN and Monte Carlo samples are very similar. There were $9834$ events for each type.}
    \label{fig:triforce_GAN_train}
\end{figure}

Triforce can also be considered as an example of a typical reconstruction tool used in HEP simulation. We test the performance of our 3DGAN generated images for training this tool. The Triforce requires two types of particles (electrons and charged pions ) for training. We train the Triforce GoogleNet model from scratch on GEANT4 electron events and then on GAN electron events. The charged pion events in both trainings are those generated by the GEANT4. Figure~\ref{fig:triforce_GAN_train} presents a comparison for the primary particle energy regression for the network trained on GEANT4 against the network trained on GAN electrons showing a similar performance. The particle type classification accuracy presented in Table~\ref{tab:class} also manifest similar values. Thus we prove that the GAN simulation can be used to replace the GEANT4 simulation without any loss of accuracy.

In the context of GAN evaluation, the GAN-train and GAN-test are two very interesting concepts~\cite{2018gan_test}. The accuracy of a classifier network trained on GAN generated events and tested on data events is termed as GAN-train. When GAN images are high quality and as diverse as the training set, the score on the validation set should be similar to training accuracy. A lower accuracy would indicate that GAN images are not covering the entire distribution of the training data. GAN-test is the accuracy of a network trained on true data and validated on GAN images. A lower accuracy would indicate that the GAN images are not sufficiently realistic while a higher accuracy could be related to mode dropping. The 3DGAN shows similar performance as the GEANT4 training data and thus the results of this test can also be regarded as proof of high accuracy and diversity of the GAN generated events.

\begin{table}[]
    \centering
    \caption{Triforce classification results}
    \begin{tabular}{p{0.3\textwidth}  p{0.35\textwidth} p{0.25\textwidth}}
         Trained on & Accuracy for MC & Accuracy for GAN \\
         \hline
         \hline
         Monte Carlo electrons & $0.9928 \pm 0.014$ & $0.9992\pm0.014$\\
         GAN electrons & $0.9884\pm0.014$ & $0.9998\pm0.014$\\
         
         \hline
         \hline
     \end{tabular}
    
    \label{tab:class}
\end{table}{}

\section{Conclusions and Future Suggestions}
\label{sec:conc}

Simulation is crucial for most HEP experiments. Monte Carlo methodology can successfully simulate particle interactions at the cost of time and resources. Fast simulation is a set of faster alternatives that have been successfully used to replace detailed simulation where some loss in accuracy can be acceptable. Recent advances in deep learning have had a tremendous impact on the HEP community and are especially interesting for simulation due to the possibility of training directly from the detector data, not possible using other methodologies. The 3DGAN is an effort aimed to simulate the detector output, as images generated by a neural network. These images are conditioned on a number of variables, having large dynamic range of values. We exploit a multi-step training process, resulting in accurate simulation for electrons, photons, and neutral pions. The accuracy for individual data features varies but is within $10\%$ of the GEANT4 simulation for all quantities. Preliminary work on charged pion simulation is also promising where the ECAL contains only partial showers, manifesting higher variance. The GAN is able to reproduce the essential features of a charged pion shower. 
Another exploratory study demonstrate the possibility of generating rare modes present in the data. We further provide an example of a practical use of the GAN-generated events. The GAN simulated events are able to train a third-party particle classification and regression tool for the correct classification of the GEANT4 events. The response of the tool trained on GAN data is similar to that trained on GEANT4 data. Finally, we would like to state that the GAN-generated showers are simulated with a speedup of three orders of magnitude. 

We would like to point out some insights from our endeavour that maybe helpful in the context of any future work. Certain domain-related features like the total deposited energy need to be hardcoded in the loss function while other features can be learned implicitly. These features include geometrical properties such as shapes and moments, level of sparsity, pixel intensity distribution, and correlation among the different features as well as the inputs. The model can learn complex distributions for individual features and the only part where the GAN struggles is reproducing the sharp drop in pixel intensities as discussed in Section~\ref{sec:results}. Apart from that, the sparse peripheral regions of the images are more difficult to be correctly generated, and there is some loss in performance for very low $E_{P}$ particles. The current training allows learning of an average response for the total deposited energy since the training relies on direct comparison for small data batches. We believe that a better formulation of the loss might result in better agreement in future efforts. The preliminary work on charged pion and rare mode simulation shows great promise. The charged pion simulation can be improved by including the HCAL data. The generation of rare modes can benefit by exploring methods like the ensembling approach, where multiple networks can be trained simultaneously for different modes present in the data. The speedup can also be further increased by exploiting parallel hardware~\cite{cray2018} that cannot yet be done for the sequential logic employed by the standard Monte Carlo tools. A distributed training~\cite{aisis2020} will be most essential for future generalization of the approach through hyperparameter scan.

\acknowledgments

This work has been conducted with the support of Intel in the framework of the CERN openlab-Intel collaboration agreement. Part of this work was conducted at  "\textit{iBanks}", the AI GPU cluster at Caltech. We acknowledge NVIDIA, SuperMicro and the Kavli Foundation for their support of "\textit{iBanks}". We thank Matt Zhang from the University of Illinois at Urbana-Champaign for help regarding the Triforce~\cite{Belayneh2019} model. 


\newpage
\appendix
\section{Hyper-parameters for 3DGAN}
\label{app:params}
Table~\ref{tab:var} presents the values for the different hyperparameters selected for 3DGAN.

\begin{table}[ht!]
    \centering
    \caption{Hyperparameters for 3DGAN fully optimized version}
    \begin{tabular}{p{0.05\textwidth}  p{0.3\textwidth} p{0.25\textwidth} p{0.35\textwidth}}
         No.& param & value & description \\
         \hline
         \hline
         1& epochs & $130$ for step $1$ and $30$ epochs for step $2$ & Number of iterations through entire training data\\
         2& batch size & $64$ & Number of samples in a minibatch\\
         3& latent size & $254$ & Size of latent vector sampled from Gaussian with mean=0 and std =1\\
         4& discriminator layers & $4$ & Convolutional layers in the discriminator\\
         5& generator layers & $7$ & Convolutional layers in the generator\\
         6& optimizer & RMSprop~\cite{rmsProp} & type of optimizer\\
         7& lr & $0.01$ & learning rate\\
         8& pre-processing cell intensities by power& $0.85$ & taking a power of cell intensities\\
         9& pre-processing target $E_p$ & scaling by $1/100$ & Dividing $E_p$ by a factor of $100$\\
         10& $W_G$ & $3$ & Weight for loss associated to real/fake probability\\
         11& $W_P$ & $0.1$ & Weight for loss associated to auxilliary energy regression task\\
         12& $W_E$ & $0.1$ & Weight for loss associated to sum of intensities\\
         13& $W_A$ & $25$ & Weight for loss associated to measured angle\\
         \hline
         \hline
     \end{tabular}
    
    \label{tab:var}
\end{table}{}


\providecommand{\href}[2]{#2}\begingroup\raggedright\endgroup

\end{document}